\documentclass[%
 reprint,
 amsmath,amssymb,
 aps,
 pra,
]{revtex4-2}

\usepackage{xcolor}
\definecolor{VibrantBlue}{HTML}{0077BB}
\definecolor{VibrantCyan}{HTML}{33BBEE}
\definecolor{VibrantTeal}{HTML}{009988}
\definecolor{VibrantOrange}{HTML}{EE7733}
\definecolor{VibrantRed}{HTML}{CC3311}
\definecolor{VibrantMagenta}{HTML}{EE3377}
\definecolor{VibrantGrey}{HTML}{BBBBBB}
\usepackage[utf8]{inputenc}
\usepackage{natbib}
\usepackage{graphicx}
\usepackage[colorlinks=true,urlcolor=VibrantBlue, citecolor=VibrantBlue, linkcolor=VibrantMagenta,anchorcolor=VibrantMagenta]{hyperref}
\usepackage{braket}
\usepackage{soul}
\usepackage{bm}
\usepackage[ruled]{algorithm2e}

\begin{document}

\title{Anomaly detection with variational quantum generative adversarial networks}
\begin{abstract}
    Generative adversarial networks (GANs) are a machine learning framework comprising a generative model for sampling from a target distribution and a discriminative model for evaluating the proximity of a sample to the target distribution. GANs exhibit strong performance in imaging or anomaly detection. However, they suffer from training instabilities, and sampling efficiency may be limited by the classical sampling procedure. We introduce variational quantum-classical Wasserstein GANs to address these issues and embed this model in a classical machine learning framework for anomaly detection. Classical Wasserstein GANs improve training stability by using a cost function better suited for gradient descent. Our model replaces the generator of Wasserstein GANs with a hybrid quantum-classical neural net and leaves the classical discriminative model unchanged. This way, high-dimensional classical data only enters the classical model and need not be prepared in a quantum circuit. We demonstrate the effectiveness of this method on a credit card fraud dataset. For this dataset our method shows performance on par with classical methods in terms of the $F_1$ score. We analyze the influence of the circuit ansatz, layer width and depth, neural net architecture parameter initialization strategy, and sampling noise on convergence and performance.
\end{abstract}
\author{Daniel Herr}
\email{danielmaherr@gmail.com}
\author{Benjamin Obert}
\author{Matthias Rosenkranz}
\altaffiliation{Current address: Cambridge Quantum Computing Limited, SW1E 6DR London, United Kingdom}
\email{matthias.rosenkranz@cambridgequantum.com}
\date{\today}
\affiliation{
 d-fine GmbH, An der Hauptwache 7, 60313 Frankfurt, Germany
}
\maketitle
\section{Introduction}
Anomaly detection is the task of identifying rare, irregular data points in a dataset, which differ significantly from the majority of samples in the dataset. This is critical in many domains including medical diagnosis~\cite{10.3389/fgene.2019.00599}, network intrusion detection~\cite{1408059}, fraud detection, industrial damage detection or monitoring sensors and detectors~\cite{tilaro2018model} (for reviews, see Refs.~\cite{chandolaAnomalyDetectionSurvey2009,pimentelReviewNoveltyDetection2014,ahmedSurveyNetworkAnomaly2016}).

A variety of methods have been proposed for anomaly detection based on probabilistic models, clustering, reconstruction, finding domain boundaries~\cite{pimentelReviewNoveltyDetection2014}, or tensor networks~\cite{TensorNetAno}. Some of these methods have been adapted to coherent quantum algorithms~\cite{PhysRevA.97.042315,PhysRevA.99.052310}. Recently, \citeauthor{AnoGAN}~\cite{AnoGAN} have proposed generative adversarial networks (GANs) as a promising classical method for anomaly detection. Their method, \emph{AnoGAN}, calculates an anomaly score for an unseen sample and can also flag the features that contribute most to the overall anomaly score. This is important for explaining the model's decision and can contribute to mitigating biases~\cite{buolamwiniGenderShadesIntersectional2018,obermeyerDissectingRacialBias2019}.

A GAN is an unsupervised machine learning framework comprising a generative model and a discriminative model. Unsupervised methods are well suited when labelling the data is very costly.  GANs use \emph{implicit} generative models. These types of models only require sampling access to the model distribution, not an explicit probability density~\cite{mohamedLearningImplicitGenerative2016}. An implicit generative model can sample efficiently from a learned model distribution. The discriminative model of the GAN learns to distinguish between generated and real data samples. Both models are parameterized by neural nets and trained concurrently. 

Quantum computers are conjectured to sample efficiently from distributions that are difficult for classical computers~\cite{aaronsonComputationalComplexityLinear2013,bremnerAveragecaseComplexityApproximate2016,boixoCharacterizingQuantumSupremacy2018}, and there is strong experimental evidence in favor of this conjecture~\cite{arute2019quantum}. This makes quantum computers -- potentially even noisy intermediate-scale quantum (NISQ) devices~\cite{preskillQuantumComputingNISQ2018} -- an interesting candidate for improving implicit generative models~\cite{du2018expressive,benedettiParameterizedQuantumCircuits2019}. Consequently, a number of authors have proposed and implemented experimentally various flavors of quantum GANs~\cite{situ2018quantum,lloyd2018quantum,HQGAN_19,zoufal2019quantum,zeng2019learning, dallaire2018quantum, hu2019quantum, huangExperimentalQuantumGenerative2020}. 

Training unsupervised methods such as GANs is often computationally hard and the learned distribution is restricted by the expressivity of the generative model. To address these issues we applied recent advances in GANs to the AnoGAN anomaly detection method and extended the resulting generative model with a hybrid quantum-classical neural net trained via a variational algorithm~\cite{cerezoVariationalQuantumAlgorithms2020,bhartiNoisyIntermediatescaleQuantum2021}.

First, to improve training stability we replace the standard GAN in the AnoGAN anomaly detection framework with a Wasserstein GAN with gradient penalty~\cite{arjovsky2017wasserstein,gulrajani2017improved}. The f-AnoGan method~\cite{fAnogan} combines a standard Wasserstein GAN with AnoGAN in the classical setting. The additional gradient penalty term in our method can improve convergence~\cite{gulrajani2017improved}. For simplicity of our quantum networks we do not use their proposed generator architecture which is inspired by autoencoders~\cite{autoencoder}. A Wasserstein GAN optimizes a different objective function than standard GANs, namely the 1-Wasserstein distance between the generated and real data distribution. Apart from theoretical advantages~\cite{arjovskyPrincipledMethodsTraining2017}, this objective also worked well in practice for our task. A number of authors have proposed extensions of the Wasserstein distance to quantum states (see e.g.~\cite{depalmaQuantumWassersteinDistance2020} and references therein). However, in this paper we use the classical 1-Wasserstein distance. Other classical approach to address training instabilities include energy-based GANs~\cite{zhaoEnergybasedGenerativeAdversarial2017} and associative adversarial networks~\cite{ariciAssociativeAdversarialNetworks2016}. The latter haven been extended to the hybrid quantum-classical setting on a quantum annealer~\cite{anschuetzNearTermQuantumClassicalAssociative2019} and a gate-based quantum computer~\cite{wilsonQuantumassistedAssociativeAdversarial2019}, where the latter did not lead to a notable stability improvement~\cite{wilsonQuantumassistedAssociativeAdversarial2019}.

Second, the hybrid quantum-classical generative model combines a parameterized quantum circuit with a classical neural net. The quantum circuit operates on a low-dimensional latent space suitable for intermediate-scale quantum devices. The classical neural net scales expectations calculated from this circuit to the higher-dimensional data space. In contrast to the quantum Wassersten GAN proposed by~\citeauthor{chakrabartiQuantumWassersteinGenerative2019a}~\cite{chakrabartiQuantumWassersteinGenerative2019a}, the discriminative model of our approach remains classical and classical data only enters the discriminative model. This circumvents the data preparation bottleneck of many quantum machine learning methods~\cite{aaronsonReadFinePrint2015}.

We tested the resulting anomaly detection method on a credit card fraud dataset~\cite{kaggleCreditFraud}. \citeauthor{killoranContinuousvariableQuantumNeural2019}~\cite{killoranContinuousvariableQuantumNeural2019} have also used this dataset to test anomaly detection with a supervised learning method for continuous-variable quantum computers. Our setup allows us to train and validate this the GAN on a moderately sized sample of the full dataset. We observe that the performance metric ($F_1$ score) of this quantum AnoGAN method is on par with the classical method using extensive simulations of a variety of architectures and system sizes. While we did not observe a clear performance boost for the quantum AnoGAN on this dataset, our architecture lends itself to other use cases, possibly using datasets from quantum mechanical processes.

This paper is organized as follows. Sec.~\ref{sec:background} describes the theoretical background of the type of Wasserstein GAN used in this paper, namely Wasserstein GAN with gradient penalty~\cite{gulrajani2017improved}, as well as the AnoGAN anomaly detection method and the credit card dataset. Section~\ref{sec:methods} introduces the quantum-classical Wasserstein GAN, its variational training algorithm and integration with AnoGAN. In Sec.~\ref{sec:results} we present simulation results of the quantum AnoGAN method with up to 9 qubits in the latent space and different architectures. We compare different quantum circuit ansätze, circuit depths, qubit counts and investigate the presence of barren plateaus. We also observe that sampling noise does not seem to influence the $F_1$ score noticeably but reduces the convergence rate of the method. For comparison we compute $F_1$ scores for different classical AnoGAN architectures.  We conclude in Sec.~\ref{sec:conclusion} and provide an outlook on future directions for anomaly detection with quantum systems.

\section{Background}\label{sec:background}
First, we describe how classical GANs are trained. We introduce the loss function for Wasserstein GANs. In our tests this loss function performed better than the naive loss function and we will use it for the quantum version as well. Afterwards, we introduce the AnoGAN architecture for anomaly detection.

\subsection{Wasserstein GAN with gradient penalty}\label{ssec:wgan}
A classifier learns to output a label $y \in \{0, 1\}$ given data $\bm x \sim p_r(\bm x)$ sampled from the data distribution $p_r(\bm x)$. Classifiers can broadly be categorized into discriminative and generative methods. Discriminative methods model the posterior $p(y|\bm x)$ and assign the most likely label given the input data, or learn a map from inputs to labels. Generative methods model the joint distribution $p(\bm x, y)$ of data and labels. For parametric models the posterior $p(y|\bm x)$ can then be calculated via the Bayes rule. More generally, the labels $y$ can be seen as latent variables given samples $\bm z \sim p(\bm z)$ of a (typically simple) distribution $p(\bm z)$~\cite{ngDiscriminativeVsGenerative2001}.

Explicit generative models describe distributions of the likelihood $p(\bm x|\bm z)$ and prior $p(\bm z)$ in parametric form, such as normal or uniform. These models are often trained using a variant of maximum likelihood estimation~\cite{murphy2012machine}. Implicit models drop the parametric requirement for the distributions. In this case, estimating the likelihood becomes intractable, and training via maximum likelihood is not possible. However, sampling from the implicit model is often easy. The training signal is typically extracted by comparing samples from the data distribution and the model distribution~\cite{mohamedLearningImplicitGenerative2016}. Training tries to push the generated samples closer to the data samples under some comparator.

\begin{figure}
    \centering
    \includegraphics[width=\columnwidth]{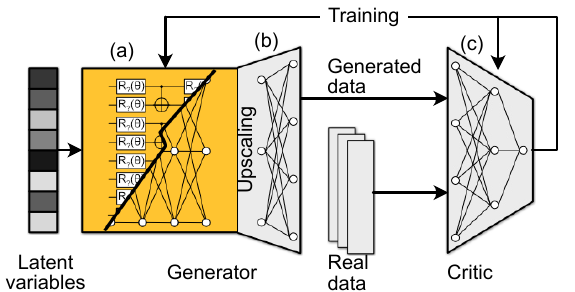}
    \caption{WGAN-GP and QWGAN-GP architectures. Given random latent variables the generator produces data whose distribution mimics the distribution of a classical dataset (`real data'). The critic determines the earth mover distance between real and generated data. The training loop adjusts the parameters of the generator and critic. The only difference between WGAN-GP and QWGAN-GP is the yellow box. For classical WGAN-GP this comprises a neural net, for QWGAN-GP this is a parameterized circuit with Pauli $Z$ measurements. Parts in grey shades are classical in both architectures.}
    \label{fig:GanArchitecture}
\end{figure}

GANs are an example of implicit generative models. Figure~\ref{fig:GanArchitecture} shows the general structure of a GAN. GANs consist of two neural nets called the \emph{generator} and \emph{discriminator} or \emph{critic}. The generator $G$ takes noise $\bm z\sim p(\bm z)$ from an easy-to-sample distribution and transforms it into a sample $G(\bm z) \sim p_g(\bm z)$ of a model distribution $p_g(\bm z)$. The discriminator $D$ outputs the probability $D(\bm x)=p(y=1|\bm x)$ of a sample $\bm x$ coming from the data distribution. Training a GAN corresponds to a min-max optimisation task. The discriminator tries to maximise the probability of assigning the correct label to an input data point while the generator tries to minimise this probability. The training signal pushes the model distribution $p_g(\bm z)$ closer to the real data distribution $p_r(\bm x)$. After successful training, the generated samples should be hard for the discriminator to distinguish from real data samples.

GANs often suffer from training instabilities due to non-convergence of the optimization, mode collapse or vanishing gradients~\cite{goodfellowGenerativeAdversarialNets2014}. To alleviate instabilities a number of authors have proposed various improvements to GANs, such as a modified cost function, architecture or regularization techniques~\cite{goodfellowGenerativeAdversarialNets2014, arjovsky2017wasserstein, salimans2016improved,diengPrescribedGenerativeAdversarial2019}. However, a large-scale comparison of popular GAN flavors showed no clear algorithmic advantage of one flavor over the other given enough computational budget for hyperparameter optimization~\cite{lucicAreGANsCreated2018}. We focus on a particular flavor called Wasserstein GAN (WGAN)~\cite{arjovsky2017wasserstein} or more specifically WGAN with gradient penalty (WGAN-GP)~\cite{gulrajani2017improved} as our tests showed better performance compared to standard GANs for our task.

A WGAN consists of a generator $G$ and a critic $D$. The critic corresponds to the discriminator in standard GANs but its output is an unbounded real number so it cannot be interpreted as a probability. The optimization task in a WGAN is
\begin{equation}
    \min_G \max_{D\in \mathcal{D}} \mathbb{E}_{\bm x \sim p_r(\bm x)}\left[D(\bm x)\right] - \mathbb{E}_{\bm z \sim p(\bm z)}\left[D\left(G(\bm z)\right)\right],
\end{equation}
where $\mathcal{D}$ is the set of $1$-Lipshitz functions $f\colon \mathbb{R}^n \rightarrow \mathbb{R}$. Recall that a function $f\colon \mathbb{R}^n\rightarrow \mathbb{R}^m$ is called 1-Lipshitz if $\lVert f(x_1) - f(x_2) \rVert \leq \lVert x_1 - x_2 \rVert$, where $\lVert\cdot\rVert$ is a metric on the respective metric spaces. To encourage the $1$-Lipshitz property during training we follow \citeauthor{gulrajani2017improved}~\cite{gulrajani2017improved} by adding a gradient penalty term to the critic loss function. This results in the WGAN-GP architecture with the following modified critic loss function
\begin{equation}\label{eq:critic_loss}
    \begin{split}
        \mathcal{L}_c(\psi) &= 
            \mathbb{E}_{p(\bm z)}\left[D_\psi\left(G(\bm z)\right)\right] - \mathbb{E}_{p_r(\bm x)}\left[D_\psi(\bm x)\right]\\
            &\quad + \lambda \mathbb{E}_{p_{\hat{\bm x}}(\hat{\bm x})}\left[\left(\left\lVert\nabla_{\hat{\bm x}} D_\psi\left(\hat{\bm x}\right)\right\rVert_2 - 1\right)^2\right],
    \end{split}
\end{equation}
where $p_{\hat{\bm x}}(\hat{\bm x})$ is the distribution of $\hat{\bm x} = \bm x + \epsilon (G(\bm z) - \bm x)$ with $\bm x \sim p_{r}(\bm x)$, $\bm z \sim p(\bm z)$, $\epsilon \sim U(0, 1)$ and $\lambda$ tunes the strength of the penalty term. Ref.~\cite{petzkaRegularizationWassersteinGANs2018} points out that this particular sampling strategy for $\hat{\bm x}$ does not strictly ensure the $1$-Lipshitz property. We do not analyze this point further as the strategy in the original paper is sufficient for training our task.

Training is commonly performed in two steps. During the first step the critic is trained on both generated and true inputs for a number of iterations by minimizing Eq.~\eqref{eq:critic_loss} with respect to the neural net parameters of the critic.  In a second step the generator is trained by minimizing the generator loss
\begin{equation}\label{eq:generator_loss}
\mathcal{L}_g(\bm{\theta}, \phi) =
    -\mathbb{E}_{p(\bm z)}\left[D\left(G(\bm z;\bm{\theta}, \phi)\right)\right].
\end{equation}
Here, $\bm{\theta}$ and $\phi$ denote parameters of the neural net of the generator, which will be detailed in Sec.~\ref{sec:methods}. The bilevel optimization task for WGAN-GP is
\begin{align}
    &\min_{\psi} \mathcal{L}_c(\psi)\\
    &\min_{\bm{\theta}, \phi} \mathcal{L}_g(\bm{\theta}, \phi).
\end{align}
After successful training, the generator should be able to produce samples from a distribution that is close to the data distribution in terms of the earth mover distance.

\subsection{AnoGAN}\label{ssec:anogan}
Anomaly detection is the task of identifying rare, irregular data points in a dataset, which differ significantly from the majority of samples in the dataset. We employ different flavors of GANs trained on the regular dataset for anomaly detection. Intuitively, one might think of feeding a new data sample to the critic and classify it as an anomaly if the critic output is below a certain threshold. In practice this approach is not sufficient because the critic is trained to distinguish generator samples from true samples and not to detect anomalous samples from the true distribution. Instead the authors of the AnoGAN architecture~\cite{AnoGAN} propose a score based on a combination of residual loss and a discrimination (here, critic) loss.  The aim is to find a latent variable $\bm z_\text{opt}$ from the otherwise uniformly sampled latent variables $z$, whose generated sample $G(\bm z_\text{opt})$ is closest to the new sample $\bm x$. The distance between them is given by the \emph{residual loss}
\begin{equation}\label{eq:residual_loss}
    \mathcal{L}_R = \left\Vert \bm x - G(\bm z_\text{opt})  \right\Vert_1.
\end{equation}
For a perfect generator, a perfect mapping to $\bm z_\text{opt}$, and a non-anomalous $\bm x$, $G(\bm z_\text{opt}) = \bm x$.

The \emph{discrimination loss} measures the distance between the optimal generated sample and $\bm x$ in an intermediate layer of the discriminator
\begin{equation}
    \mathcal{L}_D = \left\Vert f(\bm x) - f(G(\bm z_\text{opt}))  \right\Vert_1,
\end{equation}
where $f$ is some intermediate layer of the discriminator. The intuition behind this technique is that it allows comparing latent feature vectors $f$ in the discriminator instead of the only comparing scalar values $D$. However, in practice we set $f=D$, i.e. we use the scalar discriminator output as a training signal. We define the anomaly score of the AnoGAN method as
\begin{equation}
    \mathcal{S}_O = \frac{1}{\alpha} \mathcal{L}_R + \alpha \mathcal{L}_D.
    \label{eq:OutlierScore}
\end{equation}
The parameter $\alpha$ weighs the relative importance of the two losses. Given some threshold for the anomaly score we can classify new data points. We calibrate this threshold by optimizing the $F_1$ score on the test data.

One drawback of the AnoGAN architecture is its requirement for finding an optimal latent variable $\bm z_\text{opt}$ for each new input. A number of authors have proposed improvements to the classical AnoGAN architecture with the aim of reducing the computational cost of finding optimal latent variables~\cite{di2019survey,fAnogan}. The general idea of these approaches is to make the generator reversible. This is achieved using a BiGAN architecture~\cite{BiGAN, BiGANAnomalyDetection} or simultaneously training an autoencoder whose decoder is used as the generator of the GAN~\cite{ganomaly, fAnogan}. We did not investigate these methods as they increase the complexity and may result in deeper circuits or higher qubit requirements for the quantum version.

\subsection{Credit card fraud detection}\label{ssec:creditcard}
We test our models on a credit card fraud dataset downloaded from kaggle~\cite{kaggleCreditFraud}. The data were collected, cleansed, processed and labelled as part of a research collaboration between Worldline and Université Libre de Bruxelles~\cite{pozzoloCalibratingProbability2015}.

The dataset comprises 284,807 European credit card transactions over the course of 48 hours in September 2013. Features include the monetary amount of the transaction, the elapsed time from the first transaction (in seconds), and 28 covariates anonymized via a principal components analysis. Each data point is labelled either as a fraudulent or non-fraudulent transaction. We did not include the transaction time as a feature in our model so our input data contains 29 features. During a classical preprocessing step we normalize all features to the range $[0,1]$. The dataset is highly imbalanced. It contains 492 frauds out of 284,807 transactions ($0.172\%$).

\section{Methods}\label{sec:methods}
We use a WGAN-GP as the generative model for the AnoGAN method to classify credit card transactions as either fraudulent or non-fraudulent. We compare classical WGAN-GP and quantum WGAN-GP (QWGAN-GP). The QWGAN-GP uses a variational quantum-classical training algorithm.

\subsection{Classical WGAN-GP}
Figure~\ref{fig:GanArchitecture} illustrates the WGAN-GP and QWGAN-GP architectures we use in this paper. For the classical WGAN-GP, latent variables $\bm z \in \mathbb{R}^N$ are sampled from a uniform distribution $\bm z \sim U(0, 1)$. The generator comprises two parts. The first part $g_c\colon \mathbb{R}^N \rightarrow \mathbb{R}^N$ (Fig.~\ref{fig:GanArchitecture} (a)) transforms the latent variables through $L$ dense layers. Each layer is of width $N$ and has a leaky ReLU activation function with slope $0.2$~\cite{maasRectifierNonlinearitiesImprove2013}. The result of this first part is
\begin{equation}
    g_c(\bm z;\bm{\theta}_c) = (g_L \circ g_{L-1} \circ \cdots \circ g_1)(\bm z),
\end{equation}
where $\bm{\theta}_c$ are the parameters of the neural net $g_c$. We also tested other architectures for $g_c$ (cf. Fig.~\ref{fig:classical_structure}). This first part is followed by one upscaling layer (Fig.~\ref{fig:GanArchitecture} (b)) $W\colon \mathbb{R}^N \rightarrow \mathbb{R}^M$ densely connected to the output of the first part of the generator with parameters $\phi$ and a sigmoid activation function  $W(\bm x) = \sigma(w\bm x + b)$. Its purpose is to scale---typically lower-dimensional---latent variables to the data dimension $M$. For the credit card dataset $M$ is the number of features, $M=29$. The classical generator $G_c\colon \mathbb{R}^N \rightarrow \mathbb{R}^M$ is given by
\begin{equation}
    G_c(\bm z; \bm{\theta}_c, \phi) = W[g_c(\bm z; \bm{\theta}_c); \phi], \quad \bm z \sim U(0, 1).
\end{equation}

The critic is a neural net of 4 dense layers with dimensions $[M, 16, 8, 1]$ and identity activation functions, $D_\psi\colon \mathbb{R}^M \rightarrow \mathbb{R}, D_\psi = D_4\circ D_3 \circ D_2 \circ D_1$ with $D_l(\bm x; \psi_l) = d_l\bm x + b_l$ and $\psi$ is the set of all parameters of the critic. For some of our results we reduced the feature dimension to $M=N$. In this case the critic uses dense layers of dimensions $[N, 3, 3, 1]$. We do not use any batch normalization in the critic as recommended for WGAN-GP in Ref.~\cite{gulrajani2017improved}.

Training uses the Adam optimizer~\cite{kingmaAdamMethodStochastic2014}. Training of the WGAN-GP is performed on non-fraudulent data samples. Finding the optimal latent variable $\bm z_\text{opt}$ of the AnoGAN also uses Adam. The loss function for this optimization is the anomaly score Eq.~\eqref{eq:OutlierScore}. Subsequently, we determine the threshold of the anomaly score by optimizing the $F_1$ score on fraudulent and non-fraudulent data from the training set while keeping the neural net fixed. This optimization is performed with a sequential optimizer using decision trees.

\subsection{Variational quantum-classical QWGAN-GP}
\begin{figure*}
    \centering
    \includegraphics[height=3.6cm]{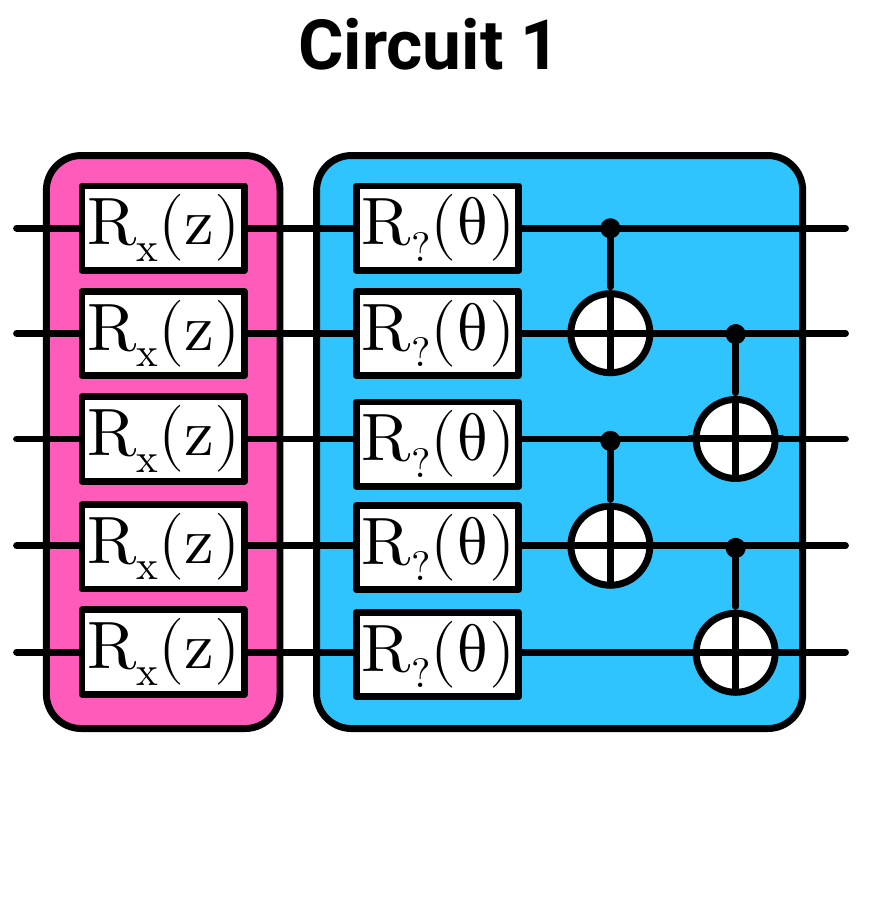}\hfill
    \includegraphics[height=3.6cm]{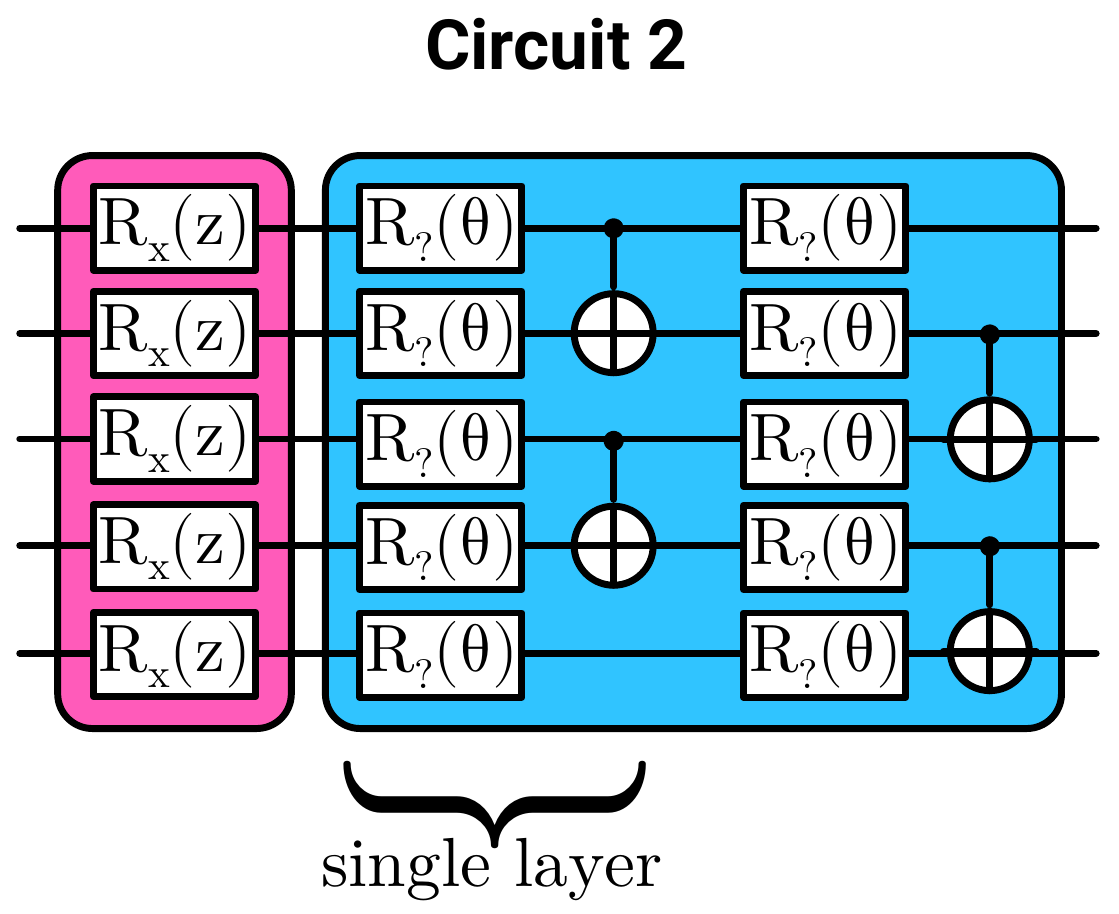}\hfill
    \includegraphics[height=3.6cm]{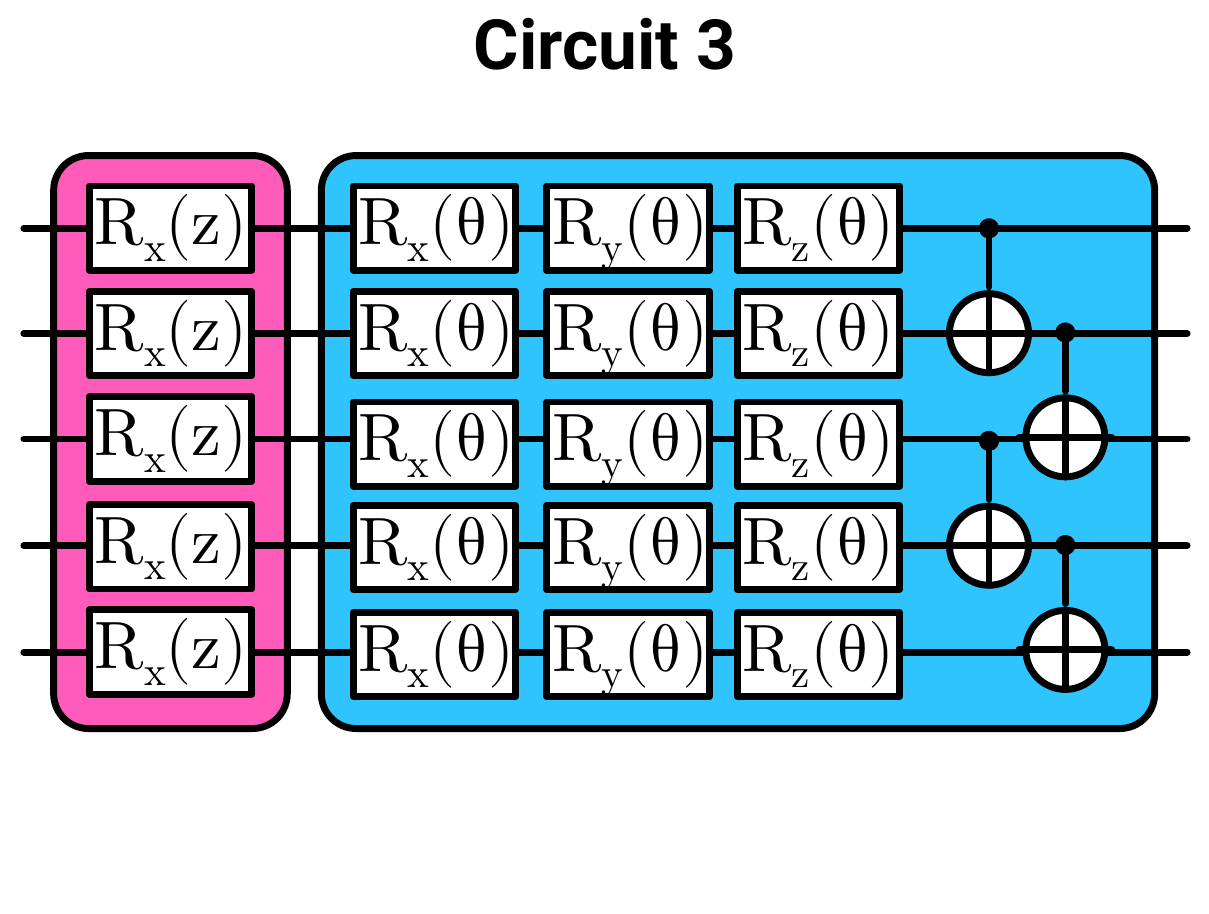}\hfill
    \includegraphics[height=3.6cm]{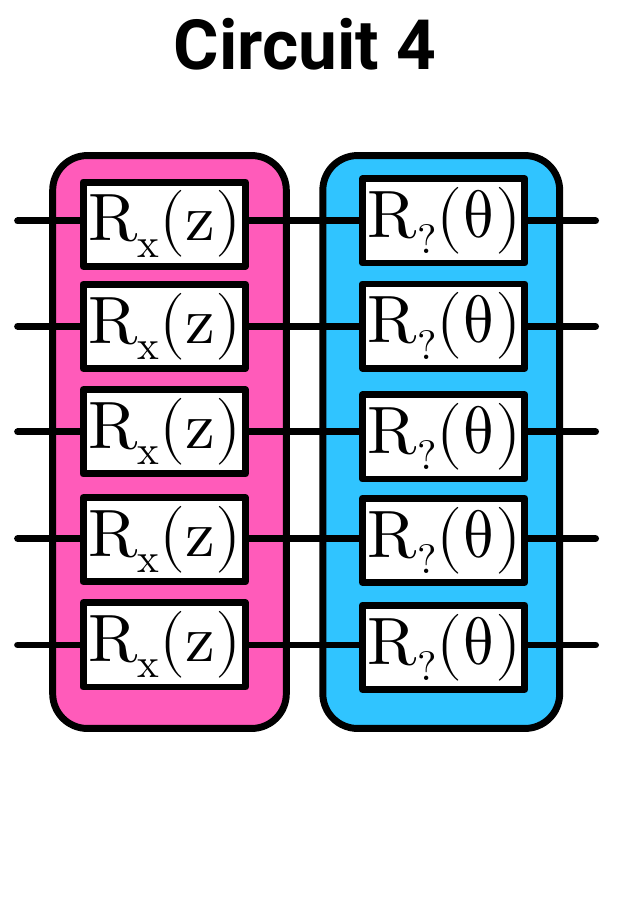}\hfill
    \caption{The parameterized circuits $U_\nu(\bm{\theta})$ analyzed in this work. These circuits together with Pauli $Z$ measurements comprise the first part of the WGAN-GP generator (Fig.~\ref{fig:GanArchitecture} (a)). The $X$ rotations in the pink box encode the latent random variable $\bm z$. The blue box is a layer of the circuit optimized by the QWGAN-GP training. Typically, a layer comprises single-qubit rotations parameterized by angles $\bm{\theta}$ and CNOT gates. The gates $R_?$ denote rotations with bases drawn uniformly from $\nu \sim \{X, Y, Z\}$ and kept fixed during a single training run. The full rotation circuit uses a fixed sequence of consecutive $X, Y, Z$ rotations.}
    \label{fig:Circuits}
\end{figure*}

For the QWGAN-GP we replace the first part of the generator from the classical WGAN-GP ($g_c$, Fig.~\ref{fig:GanArchitecture} (a)) with a short state preparation layer $S(\bm z)$, a parameterized quantum circuit $U_\nu(\bm{\theta})$ of $N$ qubits and measurements of all qubits in the $Z$ basis. The state preparation circuit comprising $N$ single-qubit $R_x$ rotations encodes $N$ uniform latent variables $\bm z \sim U(-\pi, \pi)$ as $\ket{\bm z} = S(\bm z)\ket{0}$. The classical output $g_q\colon \mathbb{R}^N \rightarrow \mathbb{R}^N$ of this first part of the QWGAN-GP generator are the $N$ expectations
\begin{equation}\label{eq:g_q}
    g_q(\bm z;\bm{\theta}, \nu) = \braket{\bm z|U_\nu^\dagger(\bm{\theta})\bm{Z} U_\nu(\bm{\theta})|\bm z}.
\end{equation}
Fig.~\ref{fig:Circuits} shows the parameterized circuit ansätze analyzed in this work. Typically, each layer of a circuit consists of single-qubit rotations with adjustable angles and CNOT gates (except for circuit 4). We analyze circuits with a variable number of these layers. The basis of each rotation is chosen uniformly random from the set $\{X, Y, Z\}$ at the beginning of a run but kept fixed during training. The parameters $\nu$ store this choice of rotation bases.

The subsequent upscaling layer $W$ (Fig.~\ref{fig:GanArchitecture} (b)) and critic (Fig.~\ref{fig:GanArchitecture} (c)) of the QWGAN-GP are identical to the classical WGAN-GP. Hence, all the differences discussed in this article are caused by variations in the generator. The overall quantum generator is given by
\begin{equation}\label{eq:quantum_generator}
    G_q(\bm z; \bm{\theta}, \phi, \nu) = W[g_q(\bm z; \bm{\theta}, \nu); \phi],\quad \bm z \sim U(-\pi, \pi).
\end{equation}

The intuitive reason for only using a quantum computer in in the generator is that quantum computers can efficiently sample from distributions which are hard to sample from classically~\cite{aaronsonComputationalComplexityLinear2013,bremnerAveragecaseComplexityApproximate2016,arute2019quantum,swekeQuantumClassicalLearnability2020}. In contrast, the critic requires loading and processing---possibly large amounts of high-dimensional---classical data, which seems unsuitable for NISQ era devices~\cite{aaronsonReadFinePrint2015}.

We train the QWGAN-GP using backpropagation through the hybrid quantum-classical net with the gradient-based Adam optimizer~\cite{bergholm2018pennylane,broughton2020tensorflow} (see Algorithm~\ref{alg:wgan} in Appendix~\ref{app:algorithms}). Gradients of the losses Eqs.~\eqref{eq:critic_loss}--\eqref{eq:generator_loss} with respect to parameter sets $\psi$ and $\phi$ are calculated with standard automatic differentiation techniques~\cite{baydinAutomaticDifferentiationMachine2018}. This means that optimizing the critic is unchanged compared to WGAN-GP but the critic loss uses samples generated by the quantum generator. Gradients $\partial \mathcal{L}_g/\partial \theta_m$ of the generator loss~\eqref{eq:generator_loss} with respect to an angle $\theta_m$ involve differentiation through an expectation of Pauli Zs for a parameterized quantum circuit. Setting $\bm\beta = \frac{\partial D}{\partial G_q} \frac{\partial G_q}{\partial g_q}$, $\bm\beta \in \mathbb{R}^N$ at fixed $\bm{\theta}$ this gradient is
\begin{equation}\label{eq:gradient_generator_loss}
    \frac{\partial\mathcal{L}_g}{\partial \theta_m}(\bm{\theta}, \phi) = -\mathbb{E}_{p(\bm z)} \left[ \bm\beta \cdot \frac{\partial}{\partial\theta_m} g_q(\bm z; \bm{\theta}, \nu) \right].
\end{equation}
We calculate the derivative of the Pauli $Z$ expectations
\begin{equation}\label{eq:gradient_g_q}
    \frac{\partial g_q}{\partial \theta_m}(\bm z; \bm{\theta}, \nu) = \frac{\partial \langle \bm Z\rangle_{\bm z; \bm{\theta}, \nu}}{\partial \theta_m}
\end{equation}
either with forward differences or the parameter shift rule~\cite{schuldEvaluatingAnalyticGradients2018,mitaraiQuantumCircuitLearning2018a} (cf. Sec.~\ref{ssec:implementation}). Note that the expectation in Eq.~\eqref{eq:gradient_generator_loss} is over the distribution of latent variables and the expectation in Eq.~\eqref{eq:gradient_g_q} is over the quantum circuit. In practice, we use a Monte Carlo estimate over a minibatch of latent variables for the former expectation and either analytical expectations or a Monte Carlo estimate over measurement samples of the state $U_\nu(\bm{\theta})\ket{\bm z}$ for the latter expectation. Note that all measurement operators commute so we only need a single measurement sample (of size $n$) for the latter estimate.

After training the QWGAN-GP on non-anomalous data, we perform anomaly detection. The first step for each anomaly detection run is finding the optimal latent variable $\bm z_\text{opt}$.

We optimize the same circuit and network as before but instead of the angles in the blue subcircuit of Fig.~\ref{fig:Circuits} we optimize the angles $\bm z$ in the state preparation layer (pink box in Fig.~\ref{fig:Circuits}). These correspond to the encoding of the latent variables. We also choose a different loss function compared to the GAN training because we want to optimize the latent variables with which the generator can produce similar data compared to the sample we want to classify. This can be done by setting the loss function to the anomaly score $\mathcal{S}_O$ from Equation~\ref{eq:OutlierScore}. We use the same Adam optimizer as we did during training of the GAN.
This procedure is identical to the optimization of latent variables in~\cite{AnoGAN}. For more information on this optimization refer to the Algorithm~\ref{alg:wgan_classification} in Appendix~\ref{app:algorithms}.

\section{Results}\label{sec:results}
\subsection{Implementation}\label{ssec:implementation}
We have implemented the classical AnoGAN with WGAN-GP in TensorFlow~\cite{tensorflow2015-whitepaper} and the quantum AnoGAN with QWGAN-GP in TensorFlow Quantum~\cite{broughton2020tensorflow}. We simulate noiseless circuits with up to 9 qubits using the default TensorFlow Quantum simulator qsim.

The results in Sec.~\ref{ssec:convergence}--\ref{ssec:latent_space} use expectations calculated from the final state of an analytical state simulation. This enables fast evaluation of the models. The results in Sec.~\ref{ssec:sampling} use expectation values from measurement samples. This allows us to judge the model's effectiveness and convergence including shot noise on more realistic devices. Gradients of analytical expectations use forward differences and gradients of sampled expectations use the parameter shift rule~\cite{schuldEvaluatingAnalyticGradients2018,mitaraiQuantumCircuitLearning2018a}.

Data points in each plot are the mean over at least 70 independent training runs. Error bands/bars are 95\% confidence intervals bootstrapped by drawing 1,000 samples from these independent runs.

\subsection{Convergence and performance}\label{ssec:convergence}
We use the $F_1$ score as a performance metric of the anomaly detection task. It is defined as
\begin{align*}
F_1 &= \frac{2}{\text{precicion}^{-1} + \text{recall}^{-1}}\\
\text{precision} &= \frac{\text{true positives}}{\text{flagged positive by classifier}}\\
\text{recall} & = \frac{\text{true positives}}{\text{actual positives}}
\end{align*}

The highly imbalanced dataset requires us to evaluate many samples for the calculation of the $F_1$ score. To alleviate some of the computational effort we sample fraudulent transactions with a higher probability. This sampling strategy has an effect on the precision and therefore the $F_1$ score. However, we apply it in both the quantum and classical AnoGAN experiments to the same extent. The results are therefore comparable. With the sampled testing set having $1/4$ fraudulent transactions, a naive discriminator that flags all transactions as fraudulent will have a precision of $1/4$ and a recall of $1$. This evaluates to a $F_1$ score of $0.4$. A worthwhile discriminator should have a higher $F_1$ score than this.

\begin{figure}
    \centering
    \includegraphics[width=\columnwidth]{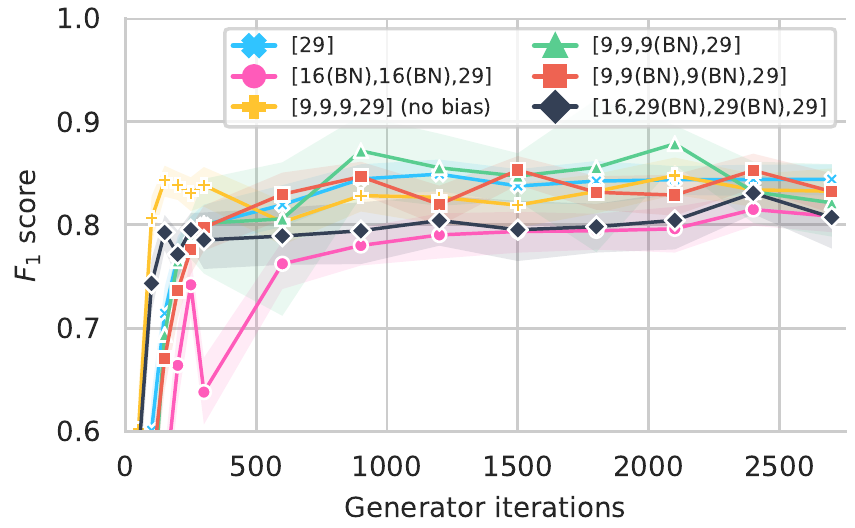}
    \caption{Convergence of AnoGANs with different classical generator architectures for latent dimension $N=9$ and critic architecture [29,16,8,1]. The numbers in square brackets are the width of each layer after the latent input layer. BN refers to batch normalisation after the respective layer. All layers in the generator are densely connected with leaky ReLU between layers and sigmoid activation after the last layer. Markers are average $F_1$ scores over 70 independent runs at a given generator training iteration and architecture, and bands are 95\% confidence intervals bootstrapped from those runs. All classical architectures perform similarly well with 95\% confidence after 2700 iterations.}\label{fig:classical_structure}
\end{figure}

As a classical baseline, Fig.~\ref{fig:classical_structure} shows the convergence of the AnoGAN $F_1$ score for several classical WGAN-GP architectures. The figure explores the impact of depth and width of the neural nets as well as batch normalization.  All classical architectures -- even the simplest one consisting of a single upscaling layer -- perform similarly well on this dataset in terms of the $F_1$ score after 2,700 training iterations. This indicates that these data pose a fairly simple anomaly detection task that does not benefit much from deeper nets.

\begin{figure}
    \centering
    \includegraphics[width=\columnwidth]{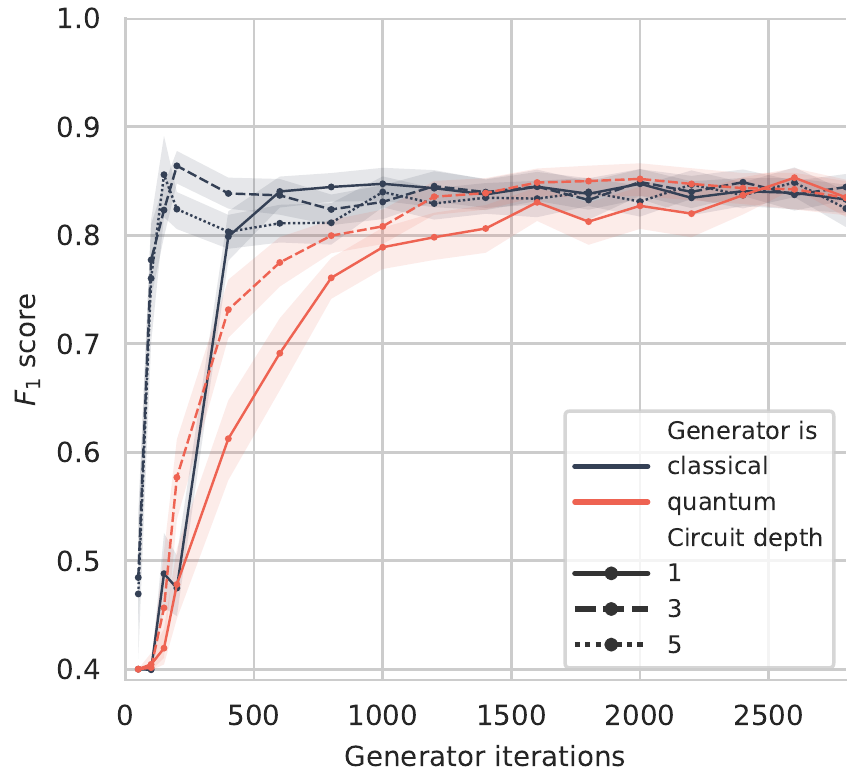}
    \caption{Convergence of $F_1$ for classical and quantum AnoGAN (with circuit 1 of Fig.~\ref{fig:Circuits}) for different depths. The latent space dimension is $9$, classical generator has constant width $9$ and the critic architecture is [29,16,8,1]. All classical and quantum architectures converge to similar mean $F_1$ scores ($F_1 \approx 0.84$).  The classical case converges in fewer iterations than the quantum case for similar depth (but higher number of parameters).}
    \label{fig:f1_vs_training}
\end{figure}

Figure~\ref{fig:f1_vs_training} compares the convergence of the AnoGAN $F_1$ scores for WGAN-GP and QWGAN-GP (circuit 1 of Fig.~\ref{fig:Circuits}) at different depths. The classical WGAN-GP uses a variable number of dense layers of width $9$ and a final dense upscaling layer of width $29$. After convergence, classical and quantum generators perform at the same level in terms of the $F_1$ score. The classical WGAN-GPs seem to converge after fewer iterations than the QWGAN-GP.  However, the number of parameters in the optimization problem is typically larger for the classical case because we use dense layers classically vs single-qubit rotations quantumly. Indeed, looking at different depths we note that deeper neural nets (quantum circuit) converge slightly faster in WGAN-GP (QWGAN-GP).  Despite the fairly tight confidence intervals overall, sometimes we have noticed a failure to converge for individual training runs. This happened both for quantum and classical AnoGAN architectures and can likely be attributed to instabilities during GAN training.

Figure~\ref{fig:quantum_structure} shows the final $F_1$ scores after 2,700 generator iterations for the different quantum circuit types of Fig.~\ref{fig:Circuits}. Circuit 3 (rotations along each axis per layer) performs best. In our simulations this comes at the cost of higher classical simulation time. For this reason, we chose circuit 1 for the remaining simulations in this paper as its performance is comparable for depths larger than 1.  Circuit 4 does not contain entangling gates. Nevertheless, its performance is also competitive for depths larger than 1. This is consistent with the earlier indication that classical neural nets are sufficient for anomaly detection with this particular dataset. The classical upscaling layer is the same as the single-layer classical generator from Fig.~\ref{fig:classical_structure}, and we can see that their $F_1$ scores are comparable. The improvement for circuit 3 with increasing depth can be explained by the randomly chosen bases of the parameterized rotations. $Z$ rotations result in changes which the $Z$ measurement cannot detect. Therefore, adding layers to the ansatz is beneficial even if there are no entangling gates.

\subsection{Initialization strategy and barren plateaus}\label{ssec:initialization}
Could barren plateaus prevent our quantum AnoGAN from gaining an edge over classical AnoGAN? Barren plateaus are flat optimization landscapes with vanishing gradients and the variance of the gradients vanishing exponentially in the number of qubits. They can prevent the gradient-based training of quantum neural networks with a randomly initialized hardware-efficient ansatz comprising local 2-designs~\cite{mccleanBarrenPlateausQuantum2018,cerezoCostFunctionDependentBarrenPlateaus2020}.   Ways to mitigate this problem are a special parameter initialization strategy using identity blocks~\cite{grant2019initialization}, layer-wise training~\cite{skolik2020layerwise} or using shallow circuits with local cost functions~\cite{cerezoCostFunctionDependentBarrenPlateaus2020}. Fortunately, our cost function is local so would not necessarily expect barren plateaus to be an issue for our numerical results with shallow circuits. Noisy simulations are not in scope of this paper so we did not study noise-induced barren plateaus~\cite{wangNoiseInducedBarrenPlateaus2020} and the signal-to-noise ratio does not suffer due to barren plateaus as it would in the noisy case~\cite{skolik2020layerwise}.

To examine further the potential impact of barren plateaus we have implemented the identity block strategy~\cite{grant2019initialization}. The idea is to start training at a point in the optimization landscape where the variance does not vanish exponentially so as to encourage early optimization steps towards the minimum. This is achieved by constructing the circuit from shallow blocks that, individually, do not approach 2-designs. The first half of each block is initialized randomly but the second half is the complex conjugate of the first (at the beginning of training). Hence, at the beginning of training the blocks evaluate to the identity. Our implementation uses a block size of 1, i.e. a depth 2 circuit contains one randomly initialized layer followed by the complex conjugate of this layer. Odd numbers of layers add one randomly initialized layer at the end. 

\begin{figure}
    \centering
    \includegraphics[width=\columnwidth]{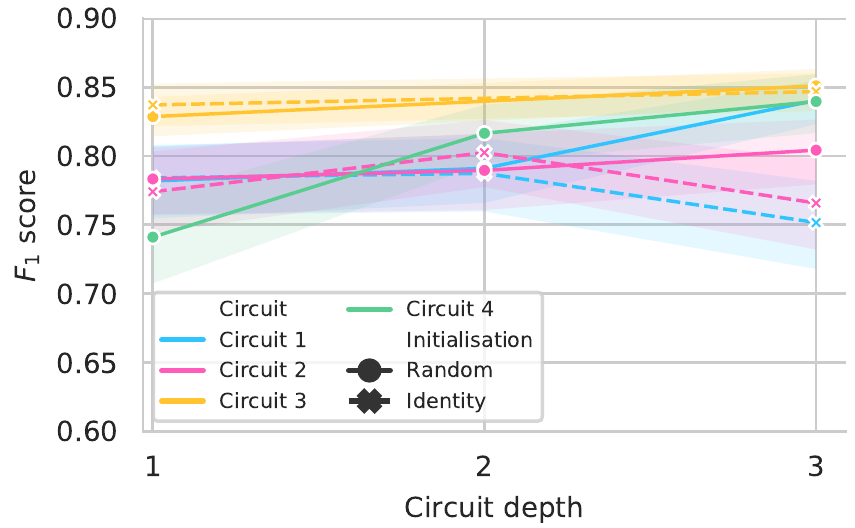}
    \caption{Average $F_1$ scores for the quantum AnoGAN with the generator circuits of Fig.~\ref{fig:Circuits} with 9 qubits and classical upscaling and classical critic architecture [29,16,8,1] as a function of circuit depth. Training was performed for 2700 generator iterations. Circuit parameters are initialised either uniformly random (solid lines) or following the identity block strategy of Ref.~\cite{grant2019initialization} (dashed). The identity initialisation performs slightly worse on average except for the full rotation circuit. This might indicate that barren plateaus are not an issue for the small qubit count and depth considered here. At depth 1 the full rotation circuit performs best (mean $F_1 \approx 0.84$) on par with the best classical architectures in Fig.~\ref{fig:classical_structure}. At depth 3 most randomly initialised circuits perform similarly well ($F_1\approx 0.85$) within 95\% confidence intervals (except circuit 2 and 1 with identity initialization).}
    \label{fig:quantum_structure}
\end{figure}

The crosses/dashed lines in Fig.~\ref{fig:quantum_structure} show the final $F_1$ scores after 2,700 generator iterations using the identity block strategy. Our anomaly detection task does not benefit from this strategy for the small system sizes studied here. Compared to the random initialization (solid lines in Fig.~\ref{fig:quantum_structure}) performance even seems to drop for some depth 3 circuits. A possible explanation is that the complex conjugate of a single layer of circuit 1 or 2 removes one layer of entangling gates compared to the random initialization strategy of the same depth.

\subsection{Dimension of the latent space}\label{ssec:latent_space}
\begin{figure}
    \centering
    \includegraphics[width=\columnwidth]{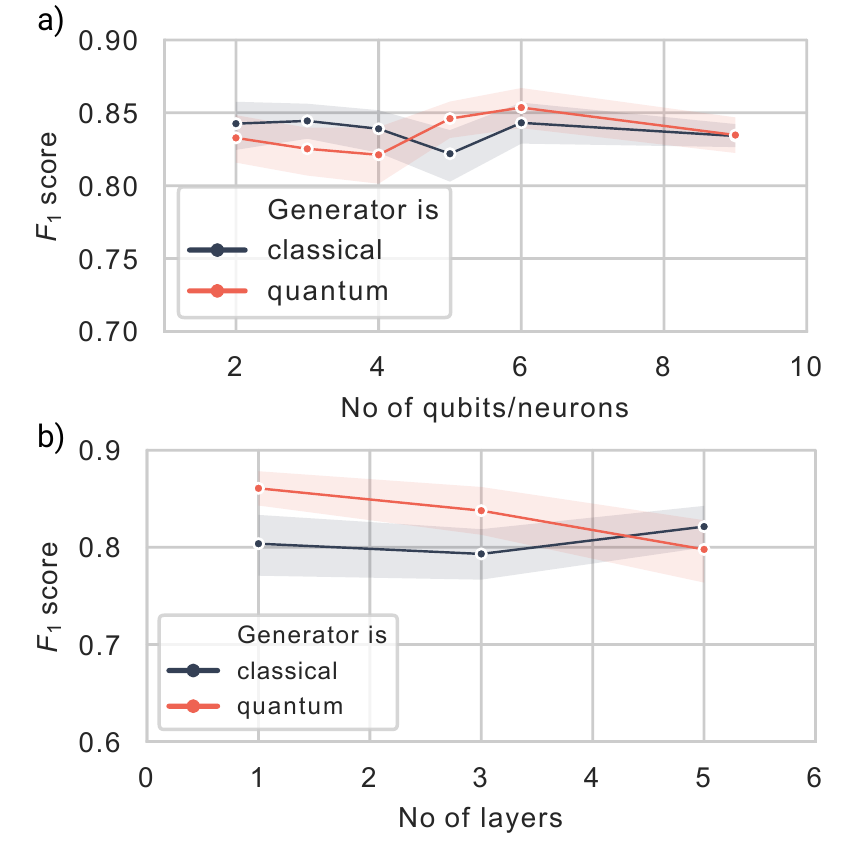}
        \caption{The effect of (a) network width and (b) network depth on the $F_1$ score. The upper plot uses the AnoGAN architecture with a (classical or quantum) generator using circuit 1 and depth 3. The bottom uses an architectures without a classical upscaling layer. Instead, the data dimension was reduced to 6 and data is reduced to the 6 most important features according to the feature importance ranking from a classical random forest output (see main text).}
    \label{fig:f1_vs_width_depth}
\end{figure}

Next we study the impact of changing the latent space dimension. This corresponds to the number of qubits in the quantum case and to the number of neurons in the classical case. Figure~\ref{fig:f1_vs_width_depth}~a) shows the final $F_1$ score after 2,700 generator iterations for different latent dimensions. The last layer in both networks is the classical upscaling layer.  We observe only a small variation of the mean final $F_1$ score. The quantum generator seems to perform slightly better than the classical generator with latent dimensions 5 and 6. However, note that the 95\% confidence intervals of quantum and classical results overlap.

The limited variation with the latent space dimension suggests that this dataset can be described by a few latent variables. A classical analysis using the random forests implementation in scikit-learn with impurity-based feature importance ranking also supports this view~\cite{breimanRandomForests2001,scikit-learn}. Here, we observe that the 3 most important features have a joint mean decrease impurity of roughly $0.4$ and the 6 most important features have a mean decrease impurity of roughly $0.6$.

Based on these observations, we restrict the features in our dataset to the 6 most important features ranked by the mean decrease impurity. This means that the real data dimension and the generator output dimension is reduced to 6. This allows us to drop the classical upscaling layer, i.e. the quantum generator consists only of the parameterized quantum circuit plus measurement. To this end Fig.~\ref{fig:f1_vs_width_depth}~b) shows the final $F_1$ scores for a quantum generator network without a classical upscaling layer and a similar classical generator network without upscaling. The quantum generator yields competitive results with only few layers despite the missing upscaling layer. The classical GAN performs slightly worse compared to the architecture with access to all the features. This behavior is not true for the quantum generator.

\subsection{Influence of sampling noise}\label{ssec:sampling}
\begin{figure}
    \centering
    \includegraphics[width=\columnwidth]{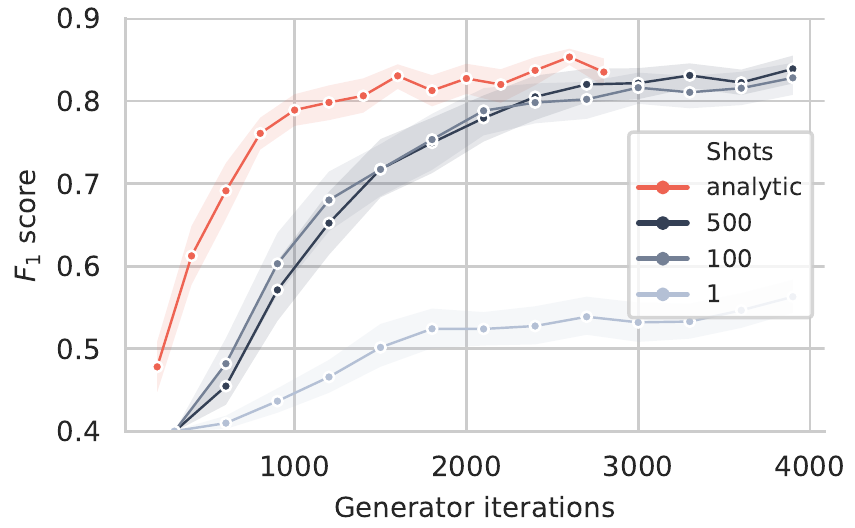}
        \caption{Convergence of the $F_1$ score for the quantum AnoGAN using analytical and sampled expectations with different sample sizes. Using sampled expectations reduces the convergence rate but does not impact the final $F_1$ score. All simulations in this plot use the same AnoGAN architecture with latent dimension 6 and a quantum generator comprising a single layer of circuit 1 and a classical upscaling layer.}
    \label{fig:sampling}
\end{figure}

On physical quantum computers, expectations are typically calculated from a finite measurement sample. In this subsection we investigate the impact of finite samples on the quantum AnoGAN performance. To this end we simulate the parameterized quantum circuit as before, sample $S$ bitstrings from the circuit, $\bm x^{(s)} \sim \lvert \braket{\bm x|U_\nu(\bm{\theta})|\bm z} \rvert^2$, and calculate expectations from this sample. In this case we calculate derivatives of the Pauli $Z$ expectations during backpropagation, Eq.~\eqref{eq:gradient_g_q}, using the parameter shift rule~\cite{schuldEvaluatingAnalyticGradients2018,mitaraiQuantumCircuitLearning2018a}
\begin{equation}
    \frac{\partial g_q}{\partial \theta_m}(\bm z; \bm\theta, \nu) = \frac{g_q(\bm z; \bm\theta + \bm\theta_m, \nu) - g_q(\bm z; \bm\theta - \bm\theta_m, \nu)}{2},
\end{equation}
where $\bm\theta_m=(\pi/2) \bm e_m$ has a single non-zero value in the $m$-th direction.

Figure~\ref{fig:sampling} compares the convergence of the $F_1$ score for analytic expectations and sampled expectations with different sample sizes. We observe that the final $F_1$ score after convergence equals the $F_1$ score with analytical expectations within the confidence intervals. Sampling does not decrease the performance of the quantum AnoGAN but increases the number of training iterations for convergence. Increasing the number of samples lets training behavior approach that of the analytical result.

\section{Conclusion}\label{sec:conclusion}
We have introduced a quantum version of the AnoGAN architecture for anomaly detection. Our method combines the outputs of a quantum-classical generative model trained with a variational algorithm and a classical discriminative model in an anomaly score. The generative model is based on a parameterized quantum circuit and a classical upscaling layer. One advantage of this setup is that high-dimensional classical data only enters the classical discriminative model and need not be encoded in a quantum state. The quantum circuit only processes data on a low-dimensional latent space. To improve training stability of the classical and quantum AnoGAN we have introduced a Wasserstein GAN instead of the standard GAN in the AnoGAN architecture.

We have compared the classical and quantum AnoGAN methods on a classical credit card fraud dataset. The quantum version reaches competitive performance in terms of the $F_1$ score. The total classical computational cost is larger for training the quantum than the classical version due to the classical simulation of the quantum circuits. However, the results of Sec.~\ref{ssec:sampling} suggest that the overall $F_1$ score would not suffer from only using a few samples per expectation evaluation on a noiseless quantum computer. We suspect that including a realistic noise model would either reduce convergence rate further or reduce the $F_1$ score. Investigating this effect would be a fruitful extension of this work.

Our results indicate that the credit card dataset is too easy for the classical AnoGAN as even very simple architectures converge to decent $F_1$ scores. This suggests that any potential higher expressivity of a parameterized quantum circuit does benefit anomaly detection for this dataset. It will be interesting to apply our quantum anomaly detection method to other datasets. Most promising is perhaps an extension to quantum datasets such as measurement data from quantum mechanical experiments or data from quantum communication protocols. The latter could lead to anomaly detection in quantum communication networks similar to the way anomaly detection is used in classical communication networks, e.g. for intrusion detection~\cite{ahmedSurveyNetworkAnomaly2016}. More generally, we also envision our quantum Wasserstein GAN component as an improved method for preparing quantum states, e.g. for loading useful distributions into a quantum state~\cite{zoufal2019quantum}.

\section*{Acknowledgements}
This work was supported by the German Federal Ministry for Economic Affairs and Energy through project PlanQK (01MK20005D).

The authors want to thank Alexandru Paler for valuable feedback.

\bibliography{references}

\newpage
\onecolumngrid

\appendix
\section{AnoGAN with QWGAN-GP algorithm}\label{app:algorithms}
Algorithm~\ref{alg:wgan} is the pseudocode for our implementation of QWGAN-GP and algorithm~\ref{alg:wgan_classification} is the pseudocode for our implementation of the AnoGAN latent variable optimization.

\begin{algorithm}
  \DontPrintSemicolon
  \SetKwInOut{Require}{Require}
  \SetKwInOut{Output}{Output}

  \Require{$U_\nu(\bm{\theta})$, a parameterized quantum circuit. $w$, the learning rate. $\beta_1$, $\beta_2$, $\varepsilon$, the Adam hyperparameters. $\lambda$, the gradient penalty weight. $m$, the minibatch size. $n_\text{critic}$, the number of iterations for the critic.}
  \BlankLine

  Initialize $\psi, \phi$ randomly, e.g. Glorot uniform\;
  Initialize $\bm{\theta} \sim U(-\pi, \pi)$ either with or without the identity block modification~\cite{grant2019initialization}\;
  Initialize rotation bases $\nu$ uniformly from $\set{X, Y, Z}$\;
  \While{$\bm{\theta}, \phi$ have not converged}{
    \For{$k = 1, \dots, n_\text{critic}$}{\label{forins}
      Sample minibatch $\set{\bm x^{(i)}}_{i=1}^m \sim p_r(\bm x)$ from the data\;
      Sample minibatch $\set{\bm z^{(i)}}_{i=1}^m \sim U(-\pi, \pi)$ from the latent distribution\;
      Sample minibatch $\set{\epsilon^{(i)}}_{i=1}^m \sim U(0,1)$\;
      
      $\bm{x}_g^{(i)} \leftarrow G_q(\bm z^{(i)}; \bm{\theta}, \phi, \nu),\quad$ for $i=1, \dots, m$\;
      $\hat{\bm x}^{(i)} \leftarrow \bm x^{(i)} + \epsilon^{(i)} \left(\bm{x}_g^{(i)} - \bm x^{(i)} \right),\quad$ for $i=1, \dots, m$\;
      $\mathcal{L}_c^{(i)} \leftarrow D_\psi\left(\bm{x}_g^{(i)}\right) - D_\psi(\bm x^{(i)})
            + \lambda \left[\left(\lVert\nabla_{\hat{\bm x}^{(i)}} D_\psi\bigl(\hat{\bm x}^{(i)}\bigr)\rVert_2 - 1\right)^2\right],\quad$ for $i=1, \dots, m$\;
      $\psi \leftarrow \text{Adam}\left(\nabla_\psi \frac{1}{m} \sum_{i=1}^m \mathcal{L}_c^{(i)}, \psi, w, \beta_1, \beta_2, \varepsilon\right)$\;
    }
    Sample minibatch $\set{\bm z^{(i)}}_{i=1}^m \sim U(-\pi, \pi)$ from the latent distribution\;
    $\mathcal{L}_g^{(i)} \leftarrow -D_\psi\left(G_q(\bm z^{(i)}; \bm{\theta}, \phi, \nu)\right),\quad$ for $i=1, \dots, m$\;
    $\phi \leftarrow \text{Adam}\left(\nabla_\phi \frac{1}{m} \sum_{i=1}^m \mathcal{L}_g^{(i)}, \phi, w, \beta_1, \beta_2, \varepsilon\right)$\;
    Calculate $\nabla_{\bm{\theta}} \mathcal{L}_g^{(i)}$ via finite differences for analytical expectations or parameter shift rule for sampled expectations\;
    $\bm{\theta} \leftarrow \text{Adam}\left(\nabla_{\bm{\theta}} \frac{1}{m} \sum_{i=1}^m \mathcal{L}_g^{(i)}, \bm{\theta}, w, \beta_1, \beta_2, \varepsilon\right)$\; 
  }
  \Output{A trained QWGAN-GP $\mathcal{N}=(G_q, D_\psi)$}
  \caption{QWGAN-GP using Adam. We use $w=0.0002$, $\beta_1=0.5$, $\beta_2=0.999$, $\varepsilon = 10^{-7}$, $\lambda=10$, $m=64$, $n_\text{critic}=5$}\label{alg:wgan}
\end{algorithm}

\begin{algorithm}
  \DontPrintSemicolon
  \SetKwInOut{Require}{Require}
  \SetKwInOut{Output}{Output}

  \Require{$\bm x$, a new input. $\alpha$, the weight between residual and discrimination loss.  $\mathcal{N}=(G_q, D_\psi)$, a trained QWGAN-GP from Algorithm~\ref{alg:wgan}. $w$, the learning rate. $\beta_1$, $\beta_2$, $\varepsilon$, the Adam hyperparameters.}
  \BlankLine
  
  Initialize $\bm z \sim U(-\pi, \pi)$\;
  \While{$\bm z$ has not converged}{
    $\bm{x}_g \leftarrow G_q(\bm z)$\;
    $\mathcal L_R \leftarrow \lVert\bm x - \bm{x}_g\rVert_1$\;
    $\mathcal L_D \leftarrow \lVert D_\psi(\bm x) - D_\psi(\bm{x}_g)\rVert_1$\;
    $\mathcal S_O \leftarrow \frac{1}{\alpha} \mathcal L_R + \alpha \mathcal L_D$\;
    $\bm z \leftarrow \text{Adam}(\nabla_{\bm z} \mathcal S_O, \bm z, w, \beta_1, \beta_2, \varepsilon)$\;
  }
  \Output{The anomaly score $\mathcal S_O(\bm x, \bm z, \alpha, \mathcal{N})$}
  \caption{AnoGAN with QWGAN-GP. Each new input $\bm x$ requires this latent variable optimization. We use $\alpha=1$ and the same learning rate and Adam hyperparameters as in Algorithm~\ref{alg:wgan}.}\label{alg:wgan_classification}
\end{algorithm}

\end{document}